\documentclass[aps,preprint]{revtex4-1} 
\usepackage[utf8]{inputenc} 
\usepackage[pdftex]{graphicx}
\usepackage{amsmath}
\usepackage{amssymb} 
\usepackage{epstopdf}
\usepackage{hyperref}

\graphicspath{{./figs/}}

\begin{document}

\title{Piecewise acceleration of electrons across a periodic solid-state structure irradiated by intense laser pulse}

\date{\today}
\author{D.~A.~Serebryakov}
\email{dms@appl.sci-nnov.ru}
\affiliation{Institute of Applied Physics of the Russian Academy of
Sciences, 46 Ulyanov St., Nizhny Novgorod 603950, Russia}
\author{I.~Yu.~Kostyukov}
\affiliation{Institute of Applied Physics of the Russian Academy of
Sciences, 46 Ulyanov St., Nizhny Novgorod 603950, Russia}
\affiliation{Lobachevsky State University of Nizhny Novgorod, 23 Gagarin Avenue, Nizhny Novgorod 603950, Russia}

\begin{abstract}
Three-dimensional particle-in-cell simulations show that the periodic solid-state structures irradiated by intense ($\sim 10^{19}$\,W/cm${}^2$) laser pulses can generate collimated electron bunches with energies up to 30 MeV (and acceleration gradient of $11.5$~GeV/cm), if the microstructure period is equal to the laser wavelength.
A one-dimensional model of piecewise acceleration in the microstructure is proposed and it is in a good agreement with the results of numerical simulations. 
It shows that the acceleration process for relativistic electrons can be theoretically infinite. 
In the simulations, the optimal target parameters (the width of the microstructure elements and the microstructure period) are determined. 
The explored parameters can be used for proof-of-principle experiments demonstrating an ultrahigh gradient acceleration by a number of identical and mutually coherent laser pulses [A.~Pukhov et~al., Eur. Phys. J. Spec. Top. 223, 1197 (2014)].
\end{abstract}

\maketitle

\section{Introduction}


The effect of strong absorption of laser energy and generation of energetic electrons from microstructured targets irradiated by intense lasers has been known for more than a decade after a number of theoretical~\cite{Margarone2012, Purvis2013, Ivanov2015, Jiang2016microengineering, Ji2017exploring, Cerchez2018enhanced} and experimental~\cite{Andreev2011, Blanco2017, Jiang2016microengineering, Ji2017exploring, Klimo2011, Ivanov2015, Jiang2014effects, Jiang2014enhancing} studies.
In particular, targets with rectangular grooves on the surface, as shown in~\cite{Klimo2011}, increase the efficiency of absorption of laser energy by approximately an order of magnitude, up to 44\% (for the laser intensity $I = 1.8\times 10^{19}$~W/cm${}^2$).
In the paper~\cite{Andreev16} the optimal microstructure dimensions for the certain class of microstructures to enhance the laser energy absorption were found for the laser pulse intensity of $10^{20}$~W/cm${}^2$.
The research is especially driven by the opportunities provided by ultrahigh contrast lasers that have become available over the recent two decades~\cite{Dromey2004,Thaury2007}, so that the microstructured target shape is not significantly damaged by the laser prepulse.
In addition, the electron accelerators based on laser-solid interaction look promising for obtaining high-charge (up to tens of nC) electron bunches, which becomes possible due to high density of solid-state materials~\cite{Zigler16,Fedeli16,Serebryakov2017}.
Usually the enhanced laser absorption is attributed to stochastic heating of the electrons in the target-vacuum interface~\cite{Sentoku2002,Sheng2002,Chopineau2019}, which leads to wide angular distribution of hot electrons.
However, the maximum energy of the electrons does not significantly exceed their oscillation energy, so that in order to acheive higher energies (i.e., $\varepsilon_{max} \gg mc^2 a_0$, where $a_0$ is the dimensionless laser amplitude given by formula $I\lambda^2 = 2.75\times10^{18}\,a_0^2$, $\lambda$ is the laser wavelength in $\mu$m, $m$ is the electron mass, $e$ is the electron charge), one needs to use different mechanisms.
One way to accelerate electrons up to very high energies is to irradiate the corrugated solid surface obliquely to resonantly excite surface plasmons which have longitudinal (with respect to the surface) field component and phase speed less than the speed on light~~\cite{Fedeli16,Cantono2018}.
This method allows one to obtain collimated electron beams with high energies, but the total distance of acceleration is always limited because the surface plasmon phase speed is always less than the speed of light.

\begin{figure}
	\centering
	\includegraphics[width=\columnwidth]{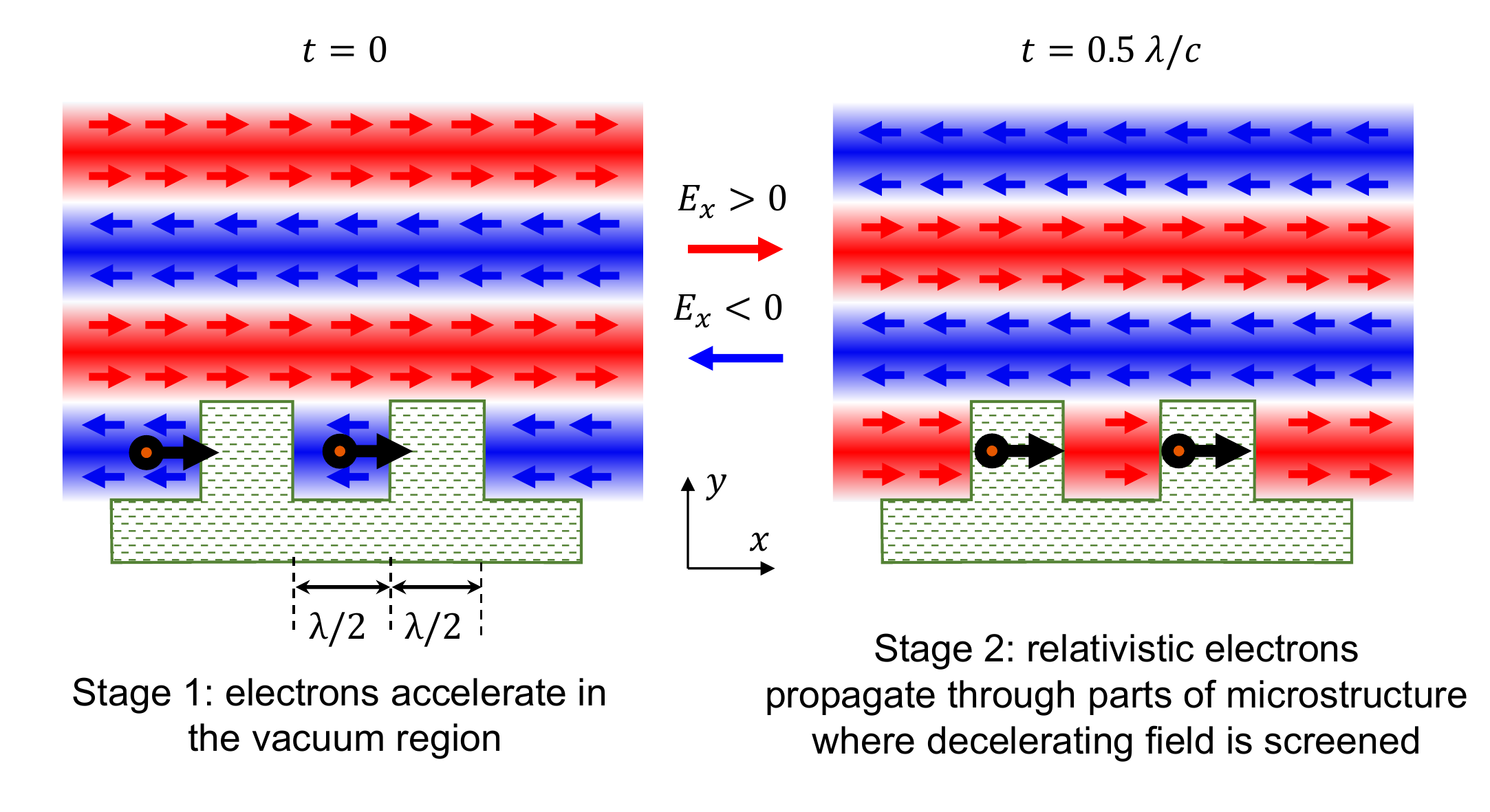}
	\caption{\label{fig:mechanism}Schematic plot demonstrating the mechanism of piecewise laser electron acceleration in the periodic structure. Each acceleration step repeat multiple times (the number of times depends on the microstructure size and the electron initial phase).}
\end{figure}

In the current paper, we study a different acceleration mechanism based on laser interaction with periodic solid-state microstructures. 
The interaction is organized in such a way that the decelerating part of the laser electric field is almost absent in the certain regions of space, so that the electron feels the laser field only in the accelerating phase for a long time.
It results in the acceleration by a piecewise-like mechanism~\cite{Pukhov2014brush, Kostyukov2015,LuuThanh2015}.
In these papers, the electrons are pre-injected into the structure and the use of phased array of fibre lasers of the ICAN project~\cite{ICAN} is suggested to achieve long acceleration distances, however, these lasers are not available yet.
Nevertheless, it turns out that the piecewise acceleration in microstructures can be achieved with conventional laser pulses as well, and without a pre-injected electron beam.
One of the goals of the paper is to propose the relatively simple proof-of-principle experimental scheme demonstrating resonance acceleration mechanism without both ICAN laser pulses and external electron beam precisely synchronized with the laser pulse.
The other goal of the paper is to get a more understandable view of the processes which occur when a relativistic laser pulse interacts with microstructured targets, and to see how it depends on the microstructure parameters.
The results may be further applied to various scenarios including high-energy particle injectors or generation of x-rays and gamma-rays~\cite{Serebryakov2019}.

The mechanism can be described as follows. 
Suppose that there is a microstructured target with rectangular elements on the surface which has the period equal to the laser wavelength, and the target is irradiated with linearly polarized relativistic laser front at normal incidence (see Fig.~\ref{fig:mechanism}).
In this case, the significant fraction of the laser front will propagate to the substrate and reflect from it, forming a standing wave between the walls of neighboring elements (in the first approximation, the ideal standing wave), and at some distance from the substrate the electric field has only the component which is parallel to the surface.
Then an electron can be extracted from the wall of the microstructure element (in a way similar to Brunel mechanism~\cite{Brunel87}) and after that accelerate in the $x$ direction to the next element (see Fig.~\ref{fig:mechanism}, left).
As the laser pulse is relativistic, the electron quickly reaches the speed close to the speed of light.
When the laser electric field changes its direction, the electron will travel approximately half of the microstructure period and will appear inside the element, where the external decelerating field is screened due to high density of solid target (see Fig.~\ref{fig:mechanism}, right).
After the full period of the laser field, the electron again appears in the space between the microstructure elements (as in Fig.~\ref{fig:mechanism}, left) in the accelerating phase of the field; and this process can be repeated many times.
It is only limited by the total length of the microstructure or the phasing conditions (however, in Sec.~\ref{sec:acc-model} it is shown that for certain initial phases the acceleration process can be theoretically infinite).
As the laser field itself is used for acceleration (not the plasma fields which are usually weaker), the acceleration gradient can be very high, which gives potential advantages over the other acceleration mechanisms.
The electron acceleration up to 200 GeV  over a distance of 10.2 cm was demonstrated by PIC simulations~\cite{Pukhov2014brush}.
The electron feels the laser field in the accelerating phase only and none in the decelerating phase due to the quickly ionized microstructures, so the acceleration process looks piece-wise. 
In some sense, the microstructures operate similar to a diode rectifying an alternating current. 
With the recent advances in micro- and nanotechnology, the fabrication of the structured targets has become a common technical process and it is possible to precisely manufacture microstructures with given period and dimensions. It increases the demand in studies of laser interaction with solid-state microstructures. 

In Sec.~\ref{sec:acc-model}, the one-dimensional analytical model of the acceleration process is presented.
It is numerically shown in Sec.~\ref{sec:simulations}, that high-energy electron bunches accelerated in the periodic structure can be obtained without an external electron beam because of efficient self-injection. The resonant nature of the process is demonstrated: the maximum electron energy drops if there is a mismatch between the laser wavelength and the microstructure period. 
The optimal width of microstructure elements is also found.
The obtained results is discussed in Sec.~\ref{sec:discussions}.

\begin{figure}
	\centering
	\def\svgwidth{0.7\columnwidth}
	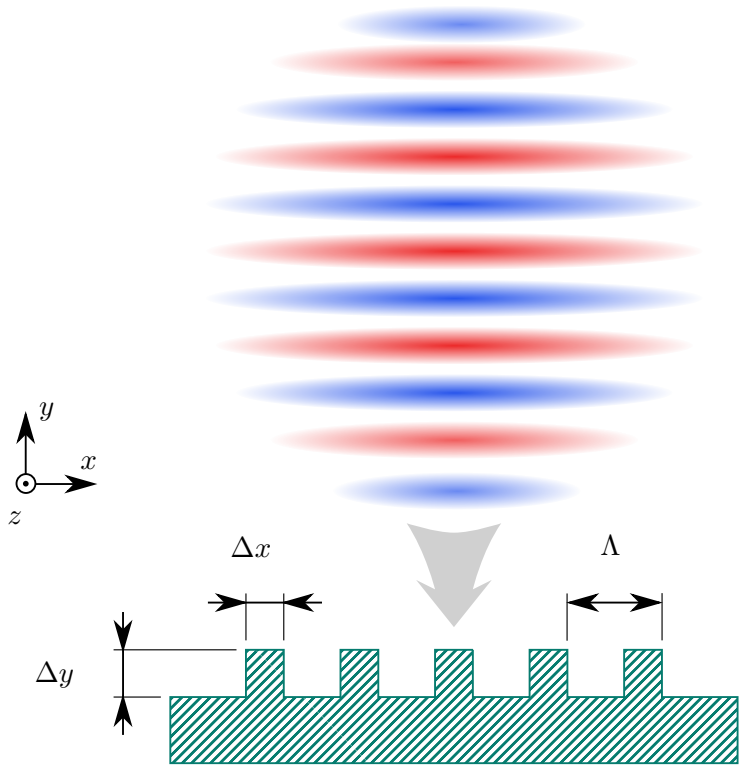
	\caption{\label{fig:schematic-microstructured}Scheme of interaction of a laser pulse with a microstructured target with rectangular-profiled elements on the surface.}
\end{figure}

\section{Electron acceleration model}
\label{sec:acc-model}

Let one consider the normal incidence of a linearly polarized laser pulse on a solid-state target with microstructured elements attached to the surface~(see Fig.~\ref {fig:schematic-microstructured}).
In the model, the width of each microstructure element $\Delta x$ and the distance between the elements are equal to half the laser wavelength $\lambda/2$ (so that the period of the structure $\Lambda$ is $\lambda$), the height of the elements $\Delta y$ is also about half of the wavelength.
The length of the the elements along the coordinate $z$ is greater than the transverse size of the pulse.
If we assume that the front of the laser pulse penetrates between the elements, maintaining a flat geometry, and ideally reflects from the substrate then the field can be described as a standing wave between the walls:
\begin{eqnarray}
E_x = -E_0\sin{k(y-y_{s})}\sin\omega t\\
B_z = E_0\cos{k(y-y_{s})}\cos\omega t,
\end{eqnarray}
where $E_0$ is the amplitude of the standing wave, $k = c/\omega = 2 \pi / \lambda$ is the laser wave number, and $y_{s}$ is the position of the surface (substrate) along the $y$ axis.
At a distance of $0.25\,\lambda$ from the surface, a node of the magnetic field of a standing wave is located, therefore, the electrons that are at a given point with a velocity directed along the axis $x$ experience only the field component $E_x$ and their dynamics can be described by one-dimensional equations:

\begin{eqnarray}
\frac{dp}{dt} & = & a_0\sin t\left[\mathrm{sign}\left(\sin x\right)+1\right],\label{eq:p-eq1}\\
\frac{dx}{dt} & = & \frac{p}{\sqrt{1+p^{2}}},\label{eq:x-eq1}
\end{eqnarray}
with the initial conditions $x(t = 0) = x_{0}$, $p(t = 0) = p_{0}$, where $\mathrm{sign}(x) = 1,0,-1$ for $x>0, \;\; x=0, \;\; x<0$, respectively, $a_0 = e E_0 / (m c \omega)$ is the dimensionless amplitude of the standing wave, $p$ is normalized  to $mc$, $t$ is normalized to $1/ \omega$, $x$ is normalized to $c/ \omega$, $e$ is the electron charge.
In the ultrarelativistic limit ($\left| p \right| \gg 1$), the equation ~(\ref{eq:x-eq1}) is reduced to a form
\begin{eqnarray}
\frac{dx}{dt} & = & 1-\frac{1}{2p^{2}},
\label{p-ur1}
\end{eqnarray}
where $\left| p \right| \gg a_0 $ is also assumed.
We suppose that the electron energy does not change when the electron moves through the material (collisions of relativistic electrons with ions at the considered laser intensities can be neglected, see the estimate~(\ref{eq:collisions})).
Writing $x_{1} = x(t = 2\pi)$, we can also neglect the change in $p$.
Then we get
\begin{eqnarray}
x_{1} & =x_{0}+2\pi & \left(1-\frac{1}{2p^{2}}\right).
\label{x-ur1}
\end{eqnarray}

Integrating Eq.~(\ref{eq:p-eq1}), in a first approximation we find

\begin{eqnarray}
p_{1} & =p_{0}+2a\, & \cos x_{0}.\label{p-ur2}
\end{eqnarray}

If we introduce $\phi_ {n} = x_{n} -2\pi n $, $ n = 1,2,3, ... $, then Eqs.~(\ref{x-ur1}) and (\ref{p-ur2}) for an arbitrary time interval $2\pi n \leq t<2 \pi \left(n + 1 \right) $
take the form of a map
\begin{eqnarray}
\label{eq:p-step-increment}
p_{n+1} & = & p_{n}+4a_0\cos\phi_{n},\label{pn-ur1}\\
\phi_{n+1} & = & \phi_{n}-\frac{\pi}{p_{n}^{2}}.\label{xn-ur1}
\end{eqnarray}
For $n \gg 1 $, the map defined by Eqs.~(\ref{x-ur1}) and (\ref{p-ur2}) can be rewritten in the form of differential equations:
\begin{eqnarray}
\frac{dp}{dt} & = & 4a_0\cos\phi,\label{dp-ur1}\\
\frac{d\phi}{dt} & = & -\frac{\pi}{p^{2}},
\label{dx-ur1}
\end{eqnarray}
which have the Hamiltonian being the integral of motion:

\begin{eqnarray}
H & = & -4a_0\sin\phi+\frac{\pi}{p}.\label{H-ur}
\end{eqnarray}
The trapped electrons correspond to infinite acceleration
($\left|p\right|\rightarrow\infty$).
The initial coordinate of electrons that can be trapped and accelerated infinitely is within the interval defined by the inequality

\begin{eqnarray}
\left|\sin\phi\right| & =\left|\frac{H\left(\left|p\right|\rightarrow\infty\right)}{4a_0}\right|=\left|\sin\phi_{0}-\frac{\pi}{4a_0 p_{0}}\right|\leq1 & .\label{trap-cond}
\end{eqnarray}

\begin{figure}
	\centering
	\includegraphics[width=\columnwidth]{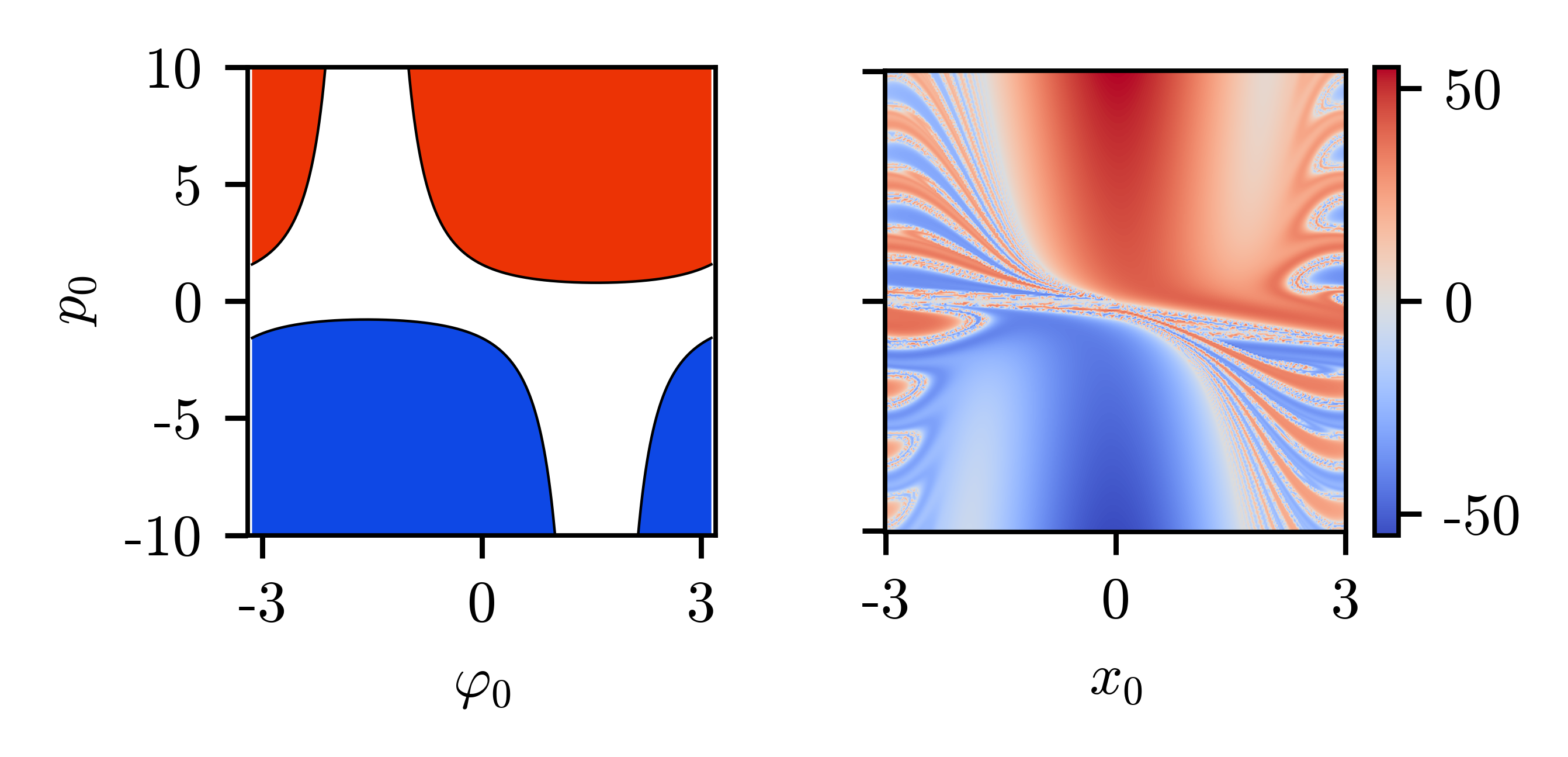}
	\caption{(Left)
	The regions of phase space corresponding to electron trapping and infinite acceleration in one-dimensional acceleration model. 
	The red color shows the region of infinite acceleration in the positive direction of the $x$ axis, the blue one corresponds to the negative direction.
	(Right) 
	The dependence of the final electron momentum on the initial momentum and the initial coordinate after 10 laser periods.
	The trajectories are taken from numerical integration of Eqs.~(\ref{eq:p-eq1}--\ref{eq:x-eq1}).}
	\label{fig:capture-area-numeric} 
\end{figure}

The region of phase space corresponding to trapping for $a_0 = 1$ is shown in Fig.~\ref{fig:capture-area-numeric}~(left).
Eqs.~(\ref{eq:p-eq1}--\ref{eq:x-eq1}) are also integrated numerically for the range of initial phases $\varphi_0$ and momenta in the range from $-10\,mc$ to $10\,mc$.
The final momentum of the electrons depending on the initial momentum and phase (that is, the coordinate) is shown in Fig.~\ref{fig:capture-area-numeric}~(right). 
It is seen from Fig.~\ref{fig:capture-area-numeric} that the numerical results is in qualitative agreement with analytical ones obtained in the ultrarelativistic approximation.
In both cases, the initial phase range corresponding to the trapping and infinite acceleration increases with an increase in the initial momentum $p_0$.

\section{Simulations}
\label{sec:simulations}

\begin{figure*}
\centering
\includegraphics[width=\textwidth]{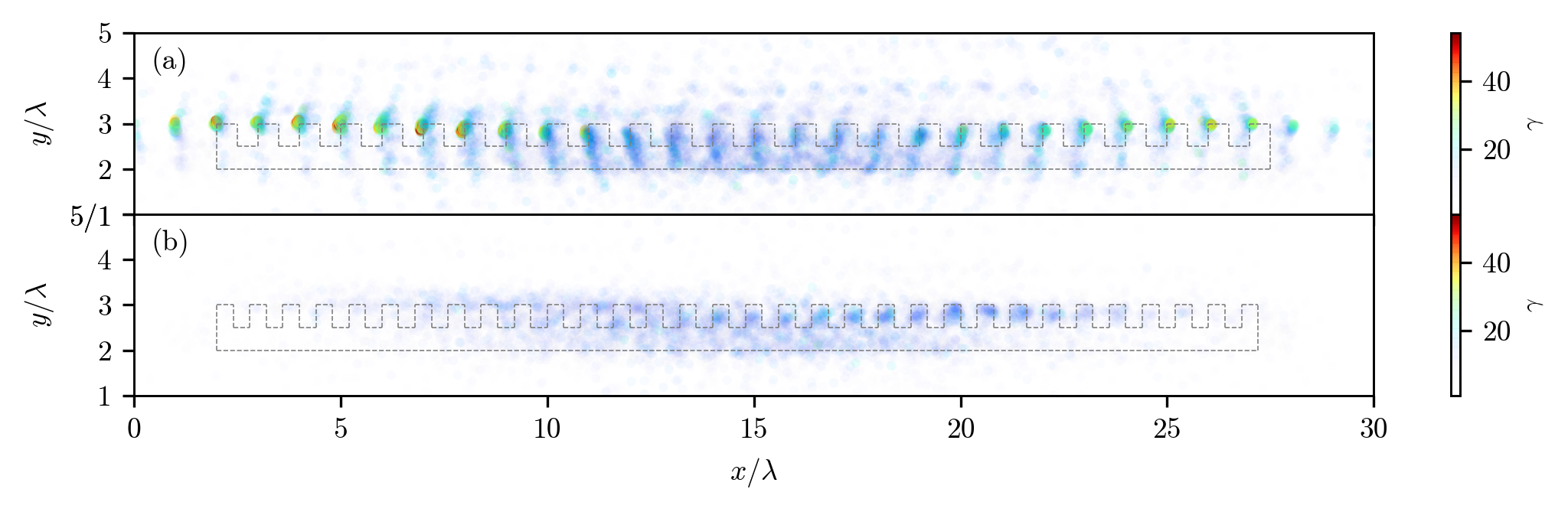}
\caption{\label{fig:spatial-dist}
The distribution of the Lorentz factor $\gamma$ of the electrons in PIC simulations for $\Delta y = 0.5\,\lambda$ and structure period $1\,\lambda$~(a) and $0.8\,\lambda$~(b), $a_0=4$.
The most energetic electrons (showed in red) are propagating predominantly in the $+x$ and $-x$ directions. 
The electrons with low $\gamma$ are hidden by the used color scheme.
The distribution is given for the end of simulations ($t=20\lambda/c$).}
\end{figure*}

The model in Sec.~\ref{sec:acc-model} has multiple assumptions, especially that the laser wavefront is not perturbed when it propagates between the microstructure elements.
In order to see the if the acceleration mechanism can be implemented in the realistic configuration, the series of three-dimensional particle-in-cell (PIC) simulations was performed.
The laser pulse in the simulations had $1~\mu$m wavelength, $1.6$~J energy and the tophat-like transverse profile with 20~$\mu$m characteristic diameter.
The pulse characteristic duration was $30$ fs, and the simulation time was $20\,\lambda/c \approx 70$~fs (around this time, the electron energy reaches its maximum value), which yielded laser intensity $I = 4.3\times 10^{19}$~W/cm${}^2$ ($a_0 = 4$).
The polarization was always linear.
We introduce the polarization angle $\varphi$ as an angle between the electric field direction and the $x$ axis (the polarization vector is assumed to be in the plane of the target). 
In most cases (unless specified explicitly), the angle $\varphi$ was equal to $0$, which means the polarization plane was perpendicular to the direction of the grooves; however, the simulation with $\varphi=\pi/2$ was also performed for comparison.
We used particle-in-cell (PIC) code \textsc{quill}~\cite{Quill,QuillVANT} which can take into account various QED effects, although they were not included into these particular simulations.
The target in the PIC simulation was represented as an array of bars with $\Delta x\times \Delta y$ size in the $xy$-plane (see Fig.~\ref{fig:schematic-microstructured}); the bars were attached to the planar plasma layer with thickness of $0.5$~$\mu$m.
The period of the microstructure $\Lambda$ was equal to $\lambda = 1~\mu$m in most cases, but in some simulations was variable.
The cross-section of the target in the simulations can be seen in Fig.~\ref{fig:spatial-dist}, where the laser pulse is incident vertically from the top side of the target.
The structure size in the $z$-direction was $24$~$\mu$m which is slightly greater than the laser pulse transverse size.
The electron density in the target was $300\,n_{cr}$ and the target was assumed to be fully ionized (the ionization process at the considered laser intensities occurs very quickly compared to the laser period), where $n_{cr} = m \omega^2 / (4 \pi e^2)$ is the critical plasma density.
The ion dynamics was not included into the simulations (the ions were considered as immobile on the simulation timescales).
The evolution of the distributions of the $E_x$ field and the electron density in the simulation with $\Delta x = \Delta y = 0.5\,\lambda$ and $\Lambda = \lambda$ is shown in Supplementary material~I.

\begin{figure}
	\includegraphics[width=\columnwidth]{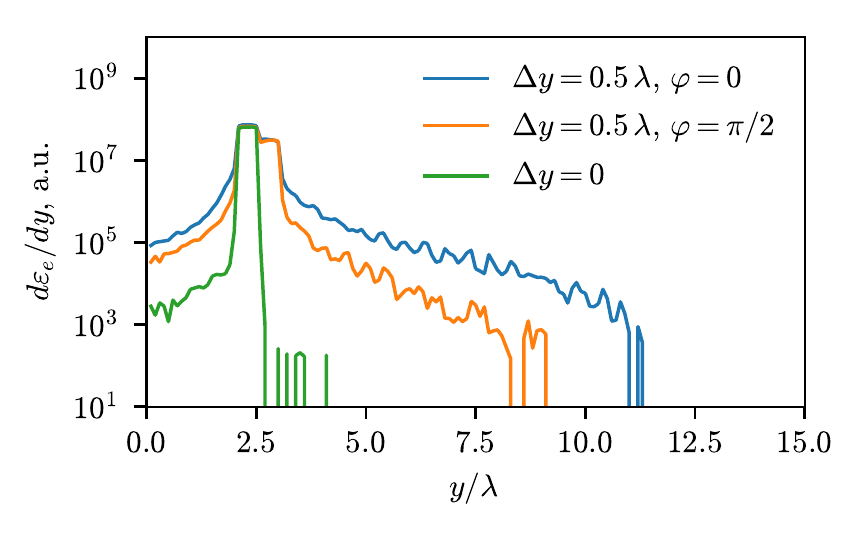}
	\caption{\label{fig:electron-cloud}The distribution of the electron energy over the coordinate $y$ which is normal to the target.
	Two cases with different laser polarization are shown: with electric field vector normal to the walls of the microstructure elements ($\varphi=0$, blue line), with electric field parallel to the walls ($\varphi=\pi/2$, orange line).
	The case of planar target is also shown for reference ($\Delta y=0$, green line).
	The angle $\varphi$ is defined as the angle between the laser pulse polarization direction and the microstructure periodicity direction (i.e., $x$ direction in Fig.~\ref{fig:schematic-microstructured}).
	The microstructure has $\Delta x =\Delta y = 0.5\,\lambda$, $\Lambda = \lambda$.
	The substrate is placed between $y=2\,\lambda$ and $y=2.5\,\lambda$, the microstructure stretches up to $3\,\lambda$.}
\end{figure}

The presence of the microstructures on the surface significantly increases the number of electrons extracted from the target, and leads to the formation of hot electron cloud above the target.
This effect highly depends on the laser polarization direction.
In Fig.~\ref{fig:electron-cloud} the distribution of the electron energy over the coordinate $y$ (i.e., the distance from the target) is shown for three numerical simulations: (i)~with microstructured target and $\varphi=0$ (blue), (ii)~with microstructured target and $\varphi=\pi/2$ (orange), (iii)~with planar target (green).
It can be seen that the polarization of the laser pulse which should be preferred for efficient generation of hot electrons corresponds to $\varphi=0$ (where the laser electric field direction is parallel to the microstructure wavevector).
This effect takes place because the electrons can be pulled from the microstructure walls by the electric field of the laser which is normal to the walls.
The formation of hot electron cloud is important for piecewise acceleration of electrons in the microstructure, because some of the electrons from the hot cloud can appear in the region between the walls with the proper momentum for acceleration according to the model from Sec.~\ref{sec:acc-model}. 
In other words, there is no need of the external electron beam because of self-injection.

The formation of high-energy electron bunches in the simulation with $\Lambda = \lambda$ and $\Delta x = \Delta y = 0.5\,\Lambda$ can be seen in Supplementary material~II, where the evolution of the spatial distribution of hot electrons in the $xy$ plane is shown.
Fig.~\ref{fig:spatial-dist}~(a) demonstrates the final snapshot of this distribution.
Color denotes the Lorentz factor of the electrons.
The high-energy electron bunches start to form at approximately $t=7\lambda/c$; they propagate through the microstructure elements and accelerate in the intervals between them.
In the $z$ direction, the electron bunches are about $15$~$\mu$m long; in some sense, they are essentially relativistic 'nanowires' with a diameter of about $100$~nm which have speed orthogonal to their direction.
The maximum Lorentz factor of the electrons is reached near the end of the simulation and equals 58, which gives the maximum electron energy of $30$~MeV.
A small asymmetry of the electron distribution is observed: the maximum energy of electrons propagating in the positive $x$ direction is lower than the maximum energy of electrons propagating in the negative direction.
This may be explained by the effect of the initial phase of the laser pulse, as the flipping the initial phase by $\pi$ (i.e., taking $\varphi=\pi$ in the simulations) results in the flipping of the electron energy distribution.
For the simulation with $\Lambda = \lambda$ and $\Delta x = \Delta y = 0.5\,\Lambda$, the total charge of the electrons with energies higher than $5$~MeV equals $11.6$~nC, with energies higher than $10$~MeV it equals $1.8$~nC.

According to Eq.~\ref{eq:p-step-increment}, the electron momentum surplus on each step can be as high as $4a_0$ (in $mc$ units), where the electric field amplitude in between the microstructure walls is assumed to be equal to $2a_0$.
In a realistic configuration, this amplitude is reduced by some factor $\alpha_f < 1$ due to non-ideal reflection from the substrate and other factors. 
So the maximum possible Lorentz factor achievable by an electron can be estimated as
\begin{equation}
\gamma_{max} = \frac{4 a_0 L}{\lambda}\alpha_{f} \alpha_{env},
\end{equation}
where $L$ is the size of the microstructure in the $x$ direction and $\alpha_{env} < 1$ is the factor that describes the longitudinal and transverse envelopes of the laser pulse (the electric field acting on the electron becomes weaker at the edges of the pulse).
For the laser pulse with both transverse and longitudinal amplitude profile $E\sim\cos^2 x$, $\alpha_{env}$ equals $0.25$, which can be used as a first-order estimate.
If the structure size $L=26\,\lambda$ as in Fig.~\ref{fig:spatial-dist}, one obtains $\gamma_{max} = 104\,\alpha_f$.
Substituting $\gamma_{max} = 58$ from the numerical results, we get $\alpha_f = 0.56$.
The average acceleration gradient in this case is $30$~MeV~/~$26$~$\mu$m = $11.5$~GeV/cm.
This is more than an order of magnitude higher than typical acceleration gradients in laser-plasma accelerators~\cite{Gonsalves2019}.

\subsection{Dependence on microstructure period}
The main characteristic feature of the piecewise acceleration mechanism is sharp dependence of the electron energy on the microstructure period, and this dependence is essentially resonant-like.
We have performed multiple numerical simulations with different periods in order to study this effect.
The hot electron spatial distribution for the two cases is shown in Fig.~\ref{fig:spatial-dist}: the so-called 'resonant' and 'non-resonant' cases, for which $\Lambda = \lambda$ and $\Lambda=0.8\,\lambda$, respectively.
In these simulations, the height of each microstructure element and its width are equal to the half of the laser wavelength $\Delta y= \Delta x = 0.5\,\lambda$.
The distribution of the electrons is analyzed at the time instant $t = 20\,\lambda/c$.
In the resonant case, it can be seen from Fig.~\ref{fig:spatial-dist} that the energetic electrons are propagating in the $+x$ and $-x$ directions in bunches separated by $\lambda$ distance due to the phase conditions.
$\gamma_{max}$ observed in the simulation is about $47$ (see Fig.~\ref{fig:spatial-dist}~(a)), which gives the maximum energy of $24$~MeV.
It should be noted that only a fraction of all particles is used for the plot, so some of the electrons in the simulations may have even higher energies which is further seen in the spectra.
The electrons on the left and right side of the target have the highest energy because they have been trapped at the time at which the laser pulse front reached the target, and they have been accelerating across the whole target.

If the microstructure period does not match the laser wavelength, the hot electron bunches are not observed  (see Fig.~\ref{fig:spatial-dist}~(b)).
It can be seen that in the non-resonant case, the hot electron energy is mostly concentrated around the center of the laser pulse, while in the resonant case the most energetic electrons are found at the edges.
This suggests that the electrons can be accelerated according to the piecewise mechanism only in the case when the microstructure period matches the laser wavelength.

\begin{figure}
	\includegraphics[width=\columnwidth]{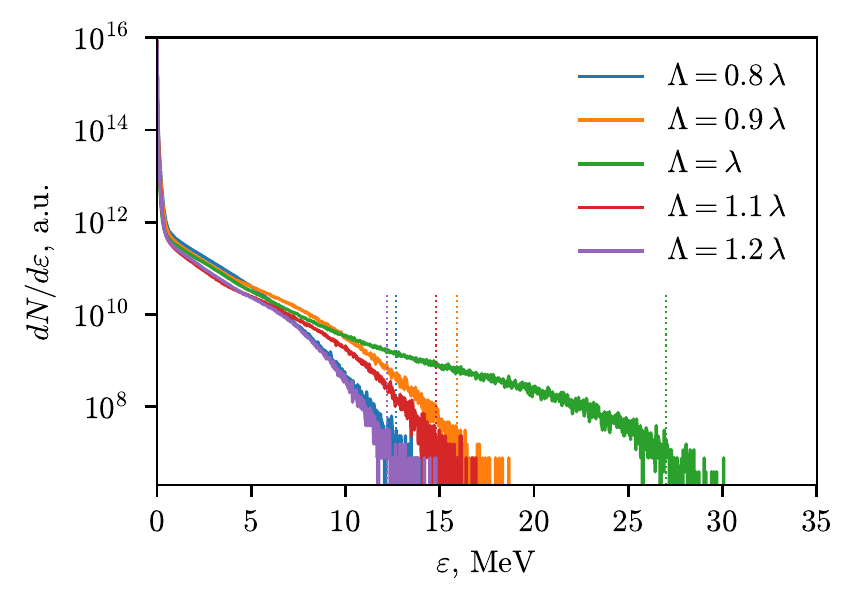}
	\caption{\label{fig:spectra}The electron spectra in simulations with different microstructure period, $a_0=4$.
		The dotted lines show the energy level $\varepsilon^\star$ for which only $10^{-7}$ fraction of all particles has higher energies.
		The microstructure has $\Delta x = 0.5\,\Lambda$ and $\Delta y = 0.5\,\lambda$.}
\end{figure}

The electron spectra for multiple simulations (the resonant and several non-resonant cases) are shown in Fig.~\ref{fig:spectra}. 
The spectra are also measured at the end of the simulation ($t = 20\,\lambda/c$).
The vertical dashed lines correspond to maximum electron energies $\varepsilon^\star$ which are computed as the maximum energy of all electrons in the simulations except for the $\eta=10^{-7}$ most energetic ones.
These values of $\varepsilon^\star$ are used further for comparison of the maximum electron energy between the simulations, instead of $\varepsilon_{max}$ (the highest electron energy in the simulation). 
Their use are more reasonable than the use of $\varepsilon_{max}$ because they are less subject to fluctuations.
The two-fold increase in the $\varepsilon^\star$ value in the resonant case ($\Lambda=\lambda$) can be seen: it equals $25.12$~MeV, and the maximum energy of individual electrons may be as high as $27.8$~MeV.
The temperature of the hot energy spectrum for the target with period $\Lambda=\lambda$ is approximately $4$~MeV (in the range between $10$ and $20$~MeV).
In comparison, the corresponding temperature in the non-resonant case ($\Lambda=0.8\,\lambda$) equals $0.95$~MeV.

\begin{figure}
	\includegraphics[width=\columnwidth]{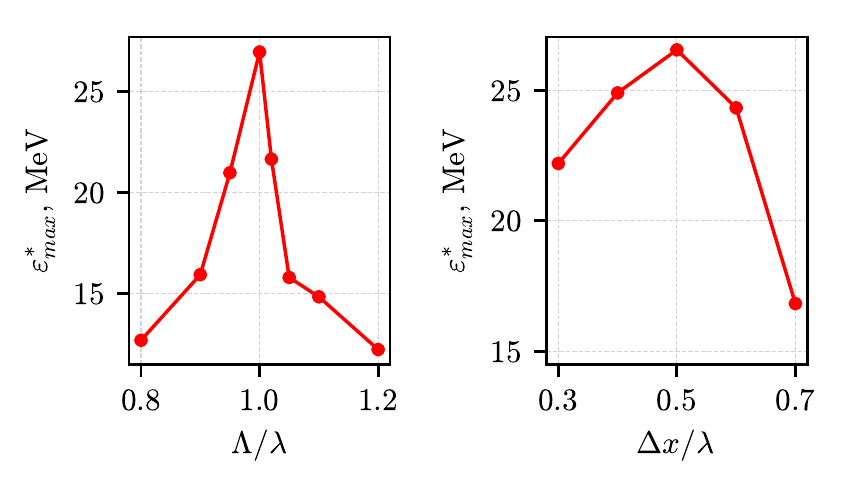}
	\caption{\label{fig:emax-period}The dependence of the maximum electron energy on the microstructure period $\Lambda$ for $\Delta x = 0.5\,\lambda$ (left) and on the width of each microstructure element $\Delta x$ for $\Lambda = \lambda$ (right).
		The energy is measured at the level of dotted lines in Fig.~\ref{fig:spectra}.
		The element height $\Delta y=0.5\,\lambda$ in all cases.}
\end{figure}

The dependence of the maximum electron energy $\varepsilon^\star$ on the microstructure period is shown in Fig.~\ref{fig:emax-period}~(left), where several additional simulations with period very close to $\lambda$ are included. The peak turns to be very sharp which supports the presence of the resonance.
In particular, the increase of the microstructure period from $\lambda$ to $1.05\,\lambda$ results in approximately $40\%$ drop in the electron energy.

\subsection{Optimal target geometry}

The optimal width of a microstructure element $\Delta x_{opt}$ is mostly determined by two factors: first, it should be close to $\lambda/2$ to screen the incident field in the improper phase, but, second, it may be more reasonable to have $\Delta x < \lambda/2$ so that the incident field can penetrate between the bars more efficiently while the effect of decelerating phase is small.
In order to determine the actual optimum, the series of simulations with different targets was performed: the width of each element varied from $0.3$ to $0.7$~$\lambda$, and other parameters were: $a_0 = 4$, $\Delta y = 0.5\,\lambda$, $\Lambda=\lambda$.
It follows from Fig.~\ref{fig:emax-period}~(right) that the target with $\Delta x_{opt} = 0.5\,\lambda$ (in this case $50\%$ of the electron trajectory is screened by the solid material) yields the maximum electron energy.
However, this peak is not so sharp as the dependence on the period: if $\Delta x = 0.4\,\lambda$ or $\Delta x=0.6\,\lambda$, the maximum electron energy decreases by only $6$--$9\%$ and is still significantly higher than in the simulations where the resonance does not occur.
If $\Delta x$ more differs from the optimal value, an asymmetry can be noted, mostly due to inefficient laser pulse penetration between the microstructure walls which are placed at the distance of only $0.3\,\lambda$. 

Comparing the results with those obtained in Ref.~\cite{Serebryakov2019} for the total absorption rate for the same microstructure period (1\,$\lambda$), one may see the following difference: the highest absorption rate (which is measured as the part of the laser energy that is contained in electrons to the end of the simulation) is achieved at $\Delta x = 0.2$--$0.3\,\lambda$ but the maximum electron energy is reached if $\Delta x = 0.5\,\lambda$.
The main reason is that the resonantly accelerated electrons comprise only the tail of the electron spectrum (as it is seen in Fig.~\ref{fig:spectra}) and their number is multiple orders of magnitude less than the number of the electrons subject to stochastic heating.
The growth of absorption rate with decrease of $\Delta x$ is also observed in the current simulations: the absorption rate equals $28\%$ for $\Delta x=0.5\,\lambda$ and $36\%$ for $\Delta x=0.3\,\lambda$.

\section{Discussions}
\label{sec:discussions}

Agreement of the results of the PIC simulations and the results of analytical model from Sec.~\ref{sec:acc-model} suggests that the electron bunches in the simulations are subject to piecewise acceleration that was analyzed in the model.
However, it is known that the effect of efficient electron acceleration in microstructured targets can be caused by different physical mechanisms, so it is important to make a distinction between them.
The electron acceleration by laser-driven relativistic surface plasmons~\cite{Fedeli16,Cantono2018} results in generation of energetic electron bunches that are propagating parallel to the corrugated surface or under a small angle to the surface, which has been demonstrated both in experiments and numerical simulations.
However, the key difference in the mechanisms is that the plasmon acceleration occurs at some distance from the grating, in the vacuum region where the surface plasma wave is excited.
Although the surface plasmons are localized near the surface (their field $E_{SP} \sim \exp(-k_{SP}y)$), the evanescent wavenumber $k_{SP}$ is about $\omega_p/\omega$ times smaller than the laser wavenumber, where $\omega_p$ is the electron plasma frequency ($\omega^2_p = 4 \pi e^2 n_e /m$).
For solid targets it means the electrons can be accelerated by surface plasmons at a distance up to multiple $\lambda$ from the target. 
However, in Fig.~\ref{fig:spatial-dist}~(a) (and in Supplementary material~II) it can be seen that the longitudinal acceleration takes place in the microstructure elements (at the distances from the substrate less than $0.5\,\lambda$) and there are no energetic electron bunches at larger distances from the target.
The energetic electrons propagate through periodic elements of the microstructure over almost the whole trajectory; only at the final stage they appear near the upper surface of the elements (at the distance $0.5\,\lambda$ from the substrate).
So, despite the hot electron cloud is formed at the distance of up to several wavelengths from the target (see Fig.~\ref{fig:electron-cloud}), the longitudinal electron acceleration in this region is not efficient as it may be expected in the case of surface plasmon acceleration.

As the numerical results from Ref.~\cite{Cantono2018} suggest, the increase in the microstructure element height from $0.25\,\lambda$ to $0.5\,\lambda$ results in more than two-fold drop in the number of hot electrons (i.e., electrons with energies above 5 MeV for $a_0=5$).
It is attributed to suppression of surface plasmon generation in the case of deep gratings.
However, in our simulations the hot electron energy increases by about 20\% if the microstructure element height grows from $0.3\,\lambda$ to $0.5\,\lambda$.
It is also an evidence that the mechanism which is responsible for acceleration in our simulation is different from acceleration by surface plasmons but is related to piecewise acceleration between the microstructure elements, as increasing the element height provides more efficient cancellation of the external field in improper phase.

Due to the complex field structure that is formed during reflection of the laser pulse from the microstructures, the electrons that are pulled into the vacuum are subject to stochastic heating~\cite{Sentoku2002,Chopineau2019}.
This mechanism is responsible for 'non-resonant' electron heating which is observed in the area above the target and the formation of hot electron cloud.
It can additionally improve self-injection because some electrons may be injected after being already accelerated until relativistic energy.

It should be noted that electron acceleration in periodic dielectric structures have been investigated over the recent years in the non-relativistic regime both theoretically~\cite{Lin2001, Huang1996, Yoder2005, Leedle15} and experimentally~\cite{Peralta2013,Breuer2013}.
The mechanism is usually referred to as DLA (dielectric laser acceleration).
It is similar to the relativistic mechanism considered in the current paper in the sense that in DLA, the accelerating field is formed due to phase effects (the laser front is propagating in dielectric medium slower that in vacuum), but the relativistic mechanism of this papers more robust since it is based on laser amplitude attenuation instead of changing of the laser phase.
It is not so strongly sensitive to the dielectric properties of the target and the height of each microstructure element (which can be difficult to control during manufacturing).
The angle of incidence in the experiment can be different from $0$ degrees, however, the laser pulse in this case needs to be $s$-polarized to keep the laser electric field parallel to the direction of microstructure periodicity.

The model that is proposed in Sec.~\ref{sec:acc-model} is purely one-dimensional and does not include the transverse electron motion.
But the transverse stability (in the $y$ direction) of the piecewise acceleration process was demonstrated in Ref.~\cite{Pukhov2014brush}: tt is shown that depending on the phase, the electrons can be either in the focusing or defocusing phase, and each region corresponds to half of the accelerating phase interval (i.e., to 25\% of the total phase interval $2\pi$).
It is seen in Fig.~\ref{fig:spatial-dist}~(a) that the electrons are concentrated in bunches in both $x$ and $y$ directions, which means that the acceleration is actually stable in transverse direction, as the leftmost electrons have propagated across the whole target with approximately constant $y$ coordinate.

Electron-ion collisions are one of the effects that should be negligible to make the described mechanism of electron acceleration achievable in experiment.
They are usually neglected for relativistic particles; however, as the electrons propagate through the target material on a relatively long distance, the collision probability increases and it should be estimated if they can be still ignored.
The collision frequency can be written as~\cite{Sentoku2008}
\begin{equation}
\nu = \frac{4\pi e^4 n_e L}{p^2 v} \approx \frac{4\pi e^4 n_e L}{m^2\gamma^2 c^3},
\end{equation}
where $L$ is the Coulomb logarithm, and $\gamma$ is the Lorentz factor of the electron.
If $n_e = 300\, n_{cr}$, one may derive	
\begin{equation}
\label{eq:collisions}
\frac{\nu}{\omega} = \frac{300 \cdot 2\pi \alpha L}{\gamma^2} \frac{\overline{\lambda_B}}{\lambda},
\end{equation}
where $\overline{\lambda_B} = \hbar/p$ is the De Broglie wavelength of an electron, $\alpha = e^2/\hbar c \approx 1/137$ is the fine-structure constant, $\hbar$ is the Planck constant.
The ratio $\overline{\lambda_B}/\lambda$ is $3.9\times 10^{-7}$ for 1~$\mu$m laser wavelength, so even for weakly relativistic electrons ($\gamma-1 \lesssim 1$) the frequency of collisions with ions is many orders of magnitude smaller than the laser frequency.

\section{Conclusions}
\label{sec:conclusions}

The simulations provided in the Sec.~\ref{sec:simulations} demonstrate the possibility to generate electron bunches with energies up to $30$~MeV and acceleration gradient of $11.5$~GeV/cm during the interaction between the intense laser pulse and the solid target with periodic microstructures.
The electron bunches are formed only if the microstructure period is close to the laser wavelength. 
The bunches in the simulations have very high total charge: $11.6$~nC for $5$~MeV threshold, and $1.8$~nC for $10$~MeV threshold.
The simulations with variable period demonstrate that the peak in the dependence is very sharp which suggests the presence of the resonant-like mechanism of acceleration.
The model of piecewise electron acceleration is studied in Sec.~\ref{sec:acc-model} which well explains the observed properties of numerical simulations (namely, the spatial distribution of the electron bunches and the optimal values of $\Delta x$).
With the help of it, it is shown that in one-dimensional approximation the electrons can be accelerated infinitely if their initial phase belongs to the certain interval.
The spatial distribution and the energy of the electron bunches in the simulations are in good agreement with the model results.

The obtained results serve multiple goals.
First, the proposed scheme is a relatively simple way to see the piecewise acceleration of electrons in experiments.
Unlike the setup proposed in Ref.~\cite{Pukhov2014brush}, the scheme does not require a pre-injected electron beam or phase-synchronized sequence of the laser pulses.
Second, the studied mechanism is an additional way to increase the efficiency of compact electron injectors based on laser-solid interaction, which require both high energy and high charge of the electron bunches.
The obtained charge of the bunches in our simulations is of the order of tens of nC, which is multiple orders of magnitude higher than usually achieved in laser-plasma electron accelerators.
Finally, the analyzed mechanism of electron acceleration can be important in other scenarios which utilize interaction of laser pulses with solid targets, including the generation of gamma-rays~\cite{Nakamura2012,Serebryakov15,Martinez2018,Serebryakov2019}.
The mechanisms that lead to generation of gamma-rays from laser-irradiated targets can be complicated, especially in the case of surface microstructures, and understanding of these mechanisms usually requires analytical and numerical modeling of electron dynamics.

\section{Acknowledgements}

This work was supported in part by the Russian Science Foundation Grant No.~18-11-00210 (PIC simulations), by the Russian Foundation For Basic Research Grants Nos.~20-52-50013 and 18-42-520054 and by the BASIS Foundation for the Advancement of Theoretical Physics and Mathematics
Grant (No. 17-11-101).
\bibliography{brush-article}
\end{document}